\documentclass[a4paper]{jpconf}
\usepackage{graphicx}
\usepackage{amsmath}
\usepackage{amssymb}
\usepackage{hyperref}
\usepackage{mathrsfs}
\usepackage{color}

\begin{document}
\title{Particle identification with the Belle II calorimeter using machine learning}
\author{Abtin Narimani Charan}

\address{Deutsches Elektronen-Synchrotron DESY, Hamburg, Germany}

\ead{abtin.narimani.charan@desy.de}

\begin{abstract}
I present an application of a convolutional neural network (CNN) to separate muons and pions in the Belle II electromagnetic calorimeter (ECL). The ECL is designed to measure the energy deposited by charged and neutral particles. It also provides important contributions to the particle identification (PID) system. Identification of low-momenta muons and pions in the ECL is crucial if they do not reach the outer muon detector. Track-seeded cluster energy images provide the maximal possible information. The shape of the energy depositions for muons and pions in the crystals around an extrapolated track at the entering point of the ECL is used together with crystal positions in $\theta-\phi$ plane and transverse momentum of the track to train a CNN. The CNN exploits the difference between the dispersed energy depositions from pion hadronic interactions and the more localized muon electromagnetic interactions. Using simulation, the performance of the CNN algorithm is compared with other PID methods at Belle II which are based on track-matched clustering information. The results show that the CNN PID method improves muon-pion separation in low momentum.

\end{abstract}

\section{Introduction}

The Belle II experiment at the SuperKEKB $e^{+} e^{-}$ collider belongs to the next generation of $B$ factories and is the upgraded version of the Belle experiment. Among the main goals of the Belle II experiment there are the search of New Physics and the high precision measurements of the Standard Model parameters in the flavour sector \cite{phys_book}. Belle II intends to collect data samples with an integrated luminosity of 50 ab$^{-1}$ at predominantly the $\Upsilon(4S)$ resonance. SuperKEKB's goal for instantaneous luminosity is 6 $\times$ 10$^{35}$ cm$^{-2}$ s$^{-1}$, which is 40 times larger than that of KEKB. This luminosity can be achieved due to a significant decrease in the beam sizes by a factor of 20 at the interaction point based on the nano-beam collision scheme and by doubling the beam currents in both rings \cite{superKEKB, tech_design}. With the above-mentioned improvements, Belle II encounters higher beam background level. The physics reach of Belle II can be enhanced if a better muon-pion separation can be achieved. A better muon-pion separation for low-momenta tracks is advantageous for $B$ physics with leptonic $\tau$ decays, since this can reduce the number of pions misidentified as muons at low momentum. It is crucial to rely on the information in the ECL because low-momentum muons (p $\lesssim$ 0.6 GeV/$c$) cannot reach the dedicated, outermost detector $K^{0}_{L}$-and-muon detector (KLM).

\section{The ECL}
\begin{figure} [htbp]
\centering
	\includegraphics[width=130mm]{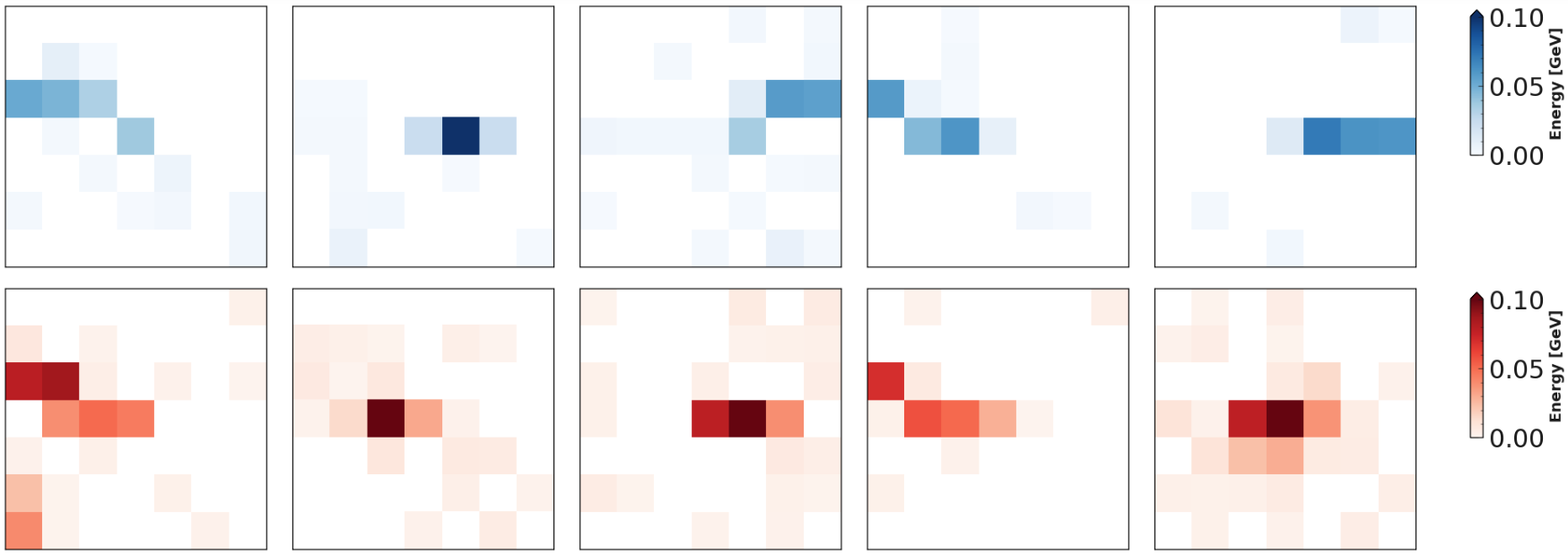}
	\caption{Patterns of energy depositions for muons (blue) and pions (red) inside the ECL barrel.}
	\label{fig:pixels}
\end{figure}

The ECL consists of 8736 thallium-doped caesium iodide CsI(Tl) crystals projected to the vicinity of the interaction point. It is immersed in a 1.5 T magnetic field and composed of a barrel and two endcaps with polar angle coverage of 12.4$^{\circ}$ -- 155.1$^{\circ}$. The dimension of each crystal is $\sim$ 5$\times$5 cm$^2$ with a length of 30 cm corresponding to 16.2 $X_{0}$ (radiation length). The barrel has 6624 crystals positioned in 46 rings of crystals distributed in the $\theta$-plane and each ring consists of 144 crystals in the $\phi$-plane \cite{phys_book, tech_design}. In this study, tracks are extrapolated into the ECL. Then, at the entry point of the track into the ECL, a window of 7$\times$7 crystals with the measured deposited energy therein is selected. Since the number of crystals in the barrel in each $\theta$-plane is equal to 144, the 7$\times$7 crystal images are symmetrical. Typical patterns of energy depositions for muons and pions are shown in figure \ref{fig:pixels}. Muons and pions both deposit energy by ionization in the matter. Additionally, pions can undergo hadronic interactions. Therefore, crystal images of muons are generally more localized than pions.

\section{Charged particle identification at Belle II}
\begin{figure} [htbp]
\centering
	\includegraphics[width=130mm]{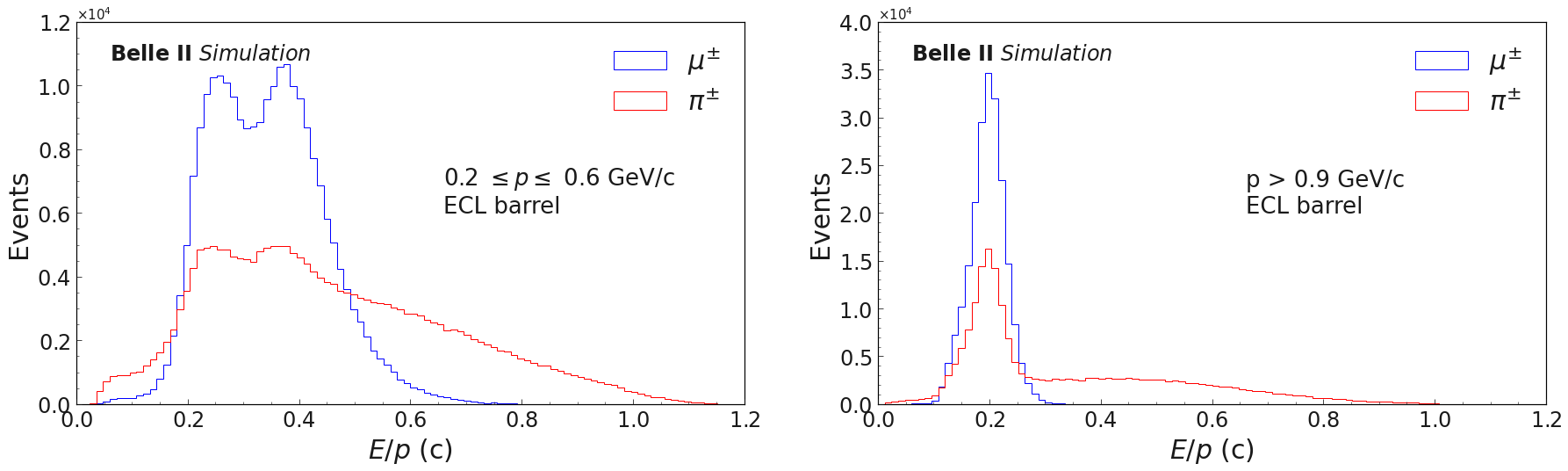}
	\caption{The distribution of the ratio of cluster energy over momentum ($E/p$) for simulated single track candidates of muons (blue) and pions (red) inside the ECL barrel for low (left) and high (right) momentum range.}
	\label{fig:Eop}
\end{figure}

There are two charged PID methods available in the Belle II analysis software framework \cite{basf2}. The first method is standard PID which is based on a combination of measurements from different sub-detectors. For each charged particle hypothesis ($i$) in each PID system, a likelihood $\mathcal{L}^{\text{PID}}_{i}$ is determined. It is a function of the probability density function parameters for a given set of observables. These likelihoods can be used to construct a combined likelihood ratio for a particular sub-detector. A binary likelihood ratio for the ECL system between muon and pion is defined as $R^{\text{ECL}}_{\mu/\pi} = \frac{\mathcal{L}^{\text{ECL}}_{\mu}}{\mathcal{L}^{\text{ECL}}_{\mu} + \mathcal{L}^{\text{ECL}}_{\pi}}$ which is used as benchmark for comparison in the study, and is indicated as \textit{default}. The standard PID in the ECL ($\mathcal{L}^{\text{ECL}}$) defines a univariate likelihood as a function of $E/p$ i.e., the ratio of reconstructed cluster energy over the momentum. The $E/p$ distribution is not very powerful for muon-pion separation specifically for low momentum tracks (figure \ref{fig:Eop}). The second method is based on boosted decision trees (BDTs) \cite{BDT}. It uses the shower-shape information in the ECL combined with likelihood information from other sub-detectors to train BDTs.

\section{Convolutional neural network (CNN)}
A convolutional neural network (CNN) is a type of deep neural network which is used to recognize visual patterns from pixel images \cite{Bishop}. In this study, two CNNs are trained with 7$\times$7 pixel images of muons and pions together with crystal positions and transverse momentum of the track in the laboratory frame (each pixel is a $\sim$ 5$\times$5 cm$^2$ crystal). Separate CNNs are trained for positive and negative charged tracks. This is due to the geometry of the ECL i.e. the direction of the crystals. Approximately 1 million single muon and pion candidates are generated with a flat distribution of transverse momentum between 0.2 and 1.0 GeV/$c$. Each track is first reconstructed in the inner tracking detectors, and then extrapolated into the ECL with Geant4 \cite{Geant4}.

\subsection{Inputs, pre-processing, and training}
There are two types of inputs for the CNN. One is the energy depositions in the 7$\times$7 pixel images and the other is a set of inputs which are fed after the convolutional layers that includes p$_{T}$, $\theta_{\text{ID}}$, and $\phi_{\text{ID}}$ of the extrapolated tracks. The $\theta_{\text{ID}}$ and $\phi_{\text{ID}}$ represent integer numbers corresponding to the location of the crystal in the ECL. The energies in the pixel images are given as they are, i.e., without any scaling applied since they already have small values. However, there are very large values in a few pixel images of pions which have energies more than 1 GeV. This is due to pion inelastic interaction with nuclei producing protons. These large values are replaced with 1 GeV. Since there are very few of these pixels, this adjustment is negligible since it affects the standard deviation and mean of the energy depositions by 0.09 \% and 0.005 \%, respectively. A threshold value of 1 MeV on the energy depositions in the pixels is applied, so that pixels below this threshold are assigned zero energy. The p$_{T}$ is already in the range of 0.2 -- 1.0 GeV/$c$, therefore no scaling is applied. The $\theta_{\text{ID}}$ and $\phi_{\text{ID}}$ are used as categorical variables which are implemented as an embedding in the network. An embedding is a distributed representation for categorical variables where each category is mapped to a distinct vector that a neural network can learn during training. The number of simulated events is identical for signal and background. The equal size of the sample is important to avoid bias against a particular type of particle or charge.

\begin{figure} [htbp]
\centering
	\includegraphics[width=120mm]{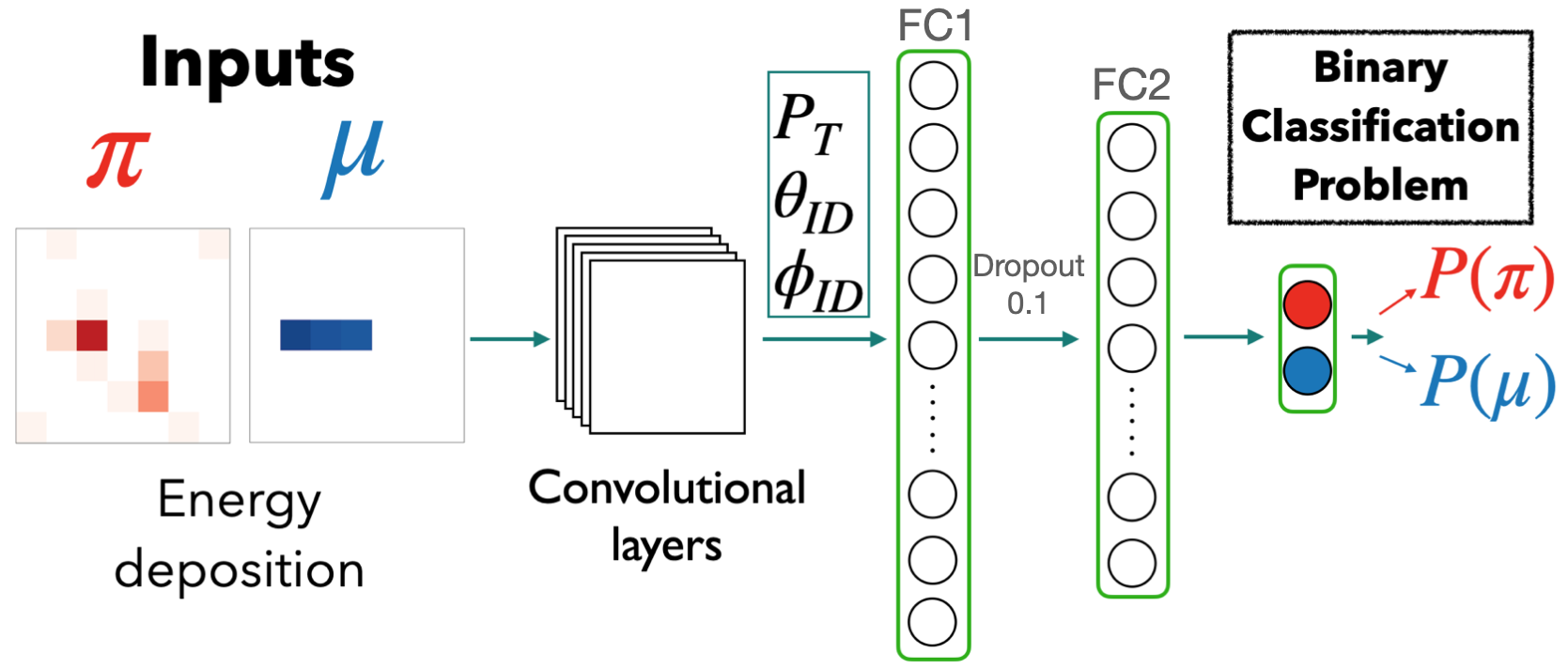}
	\caption{Neural network architecture. See text for details.}
	\label{figs:CNN_arch}
\end{figure}

The CNN includes two parts. In the first part, the 7$\times$7 pixel images of muons and pions are used as inputs for a convolutional layer. During convolution, a window of 3$\times$3, called kernel, goes through each image. A padding and stride of (1, 1) is used for the images. The padding adds one pixel with value zero on the edge of the image which is beneficial to capture more information on the edges. The stride refers to the amount of kernel movement over the image. The feature map is set to 64. The second part of the training involves a feed-forward neural network (FNN). In this part, the results from the convolutional layer must be flattened and added on top of p$_T$, $\theta_{\text{ID}}$, and $\phi_{\text{ID}}$. The number of neurons in the first and second layer of FNN are 3295 and 128, respectively. The dropout layer with a value of 0.1 is used between the first and second layer of the FNN. An \texttt{Adam} optimizer is used during the training with a learning rate of 0.001. The loss used in this study is \texttt{CrossEntropy} which is suitable for binary classification problems. The number of epochs is set to 100. The model is saved based on the lowest validation loss value in the corresponding epoch. A schematic of the network is shown in figure \ref{figs:CNN_arch}.

\subsection{Performance}
\begin{figure} [htbp]
\centering
	\includegraphics[width=130mm]{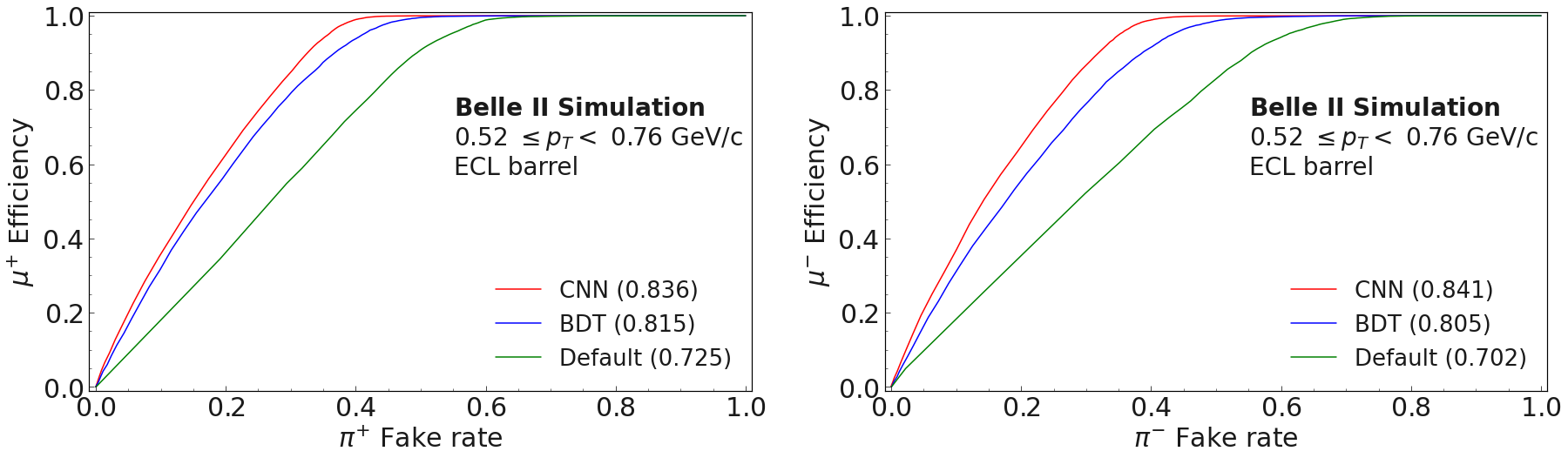}
	\caption{Comparison of CNN, BDT, and default PID. The left and right plots show positive and negative charged tracks, respectively. The value in front of each method shows the area under the curve (AUC).}
	\label{figs:roc_cnn_bdt_default}
\end{figure}

The performance of the CNN is evaluated on a test dataset with an equal number of muons and pions. The test dataset is generated and reconstructed with the exact same conditions as the training dataset. The CNN is compared with two other methods (default and BDT). The performance of all methods for both charged tracks in a range of transverse momentum (0.52 $\leq$ p$_{T}$ $<$ 0.76 GeV/$c$) are shown in figure \ref{figs:roc_cnn_bdt_default}. The $\mu$ efficiency is defined as the ratio of the number of correctly identified muons over the total number of muons. The $\pi$ fake rate is defined as the ratio of the number of pions identified as muons over the total number of pions. The CNN method outperforms the other methods with AUC (Area Under the Curve) scores of 0.836 and 0.841 for positive and negative charged tracks, respectively.

\begin{figure} [htbp]
\centering
	\includegraphics[width=130mm]{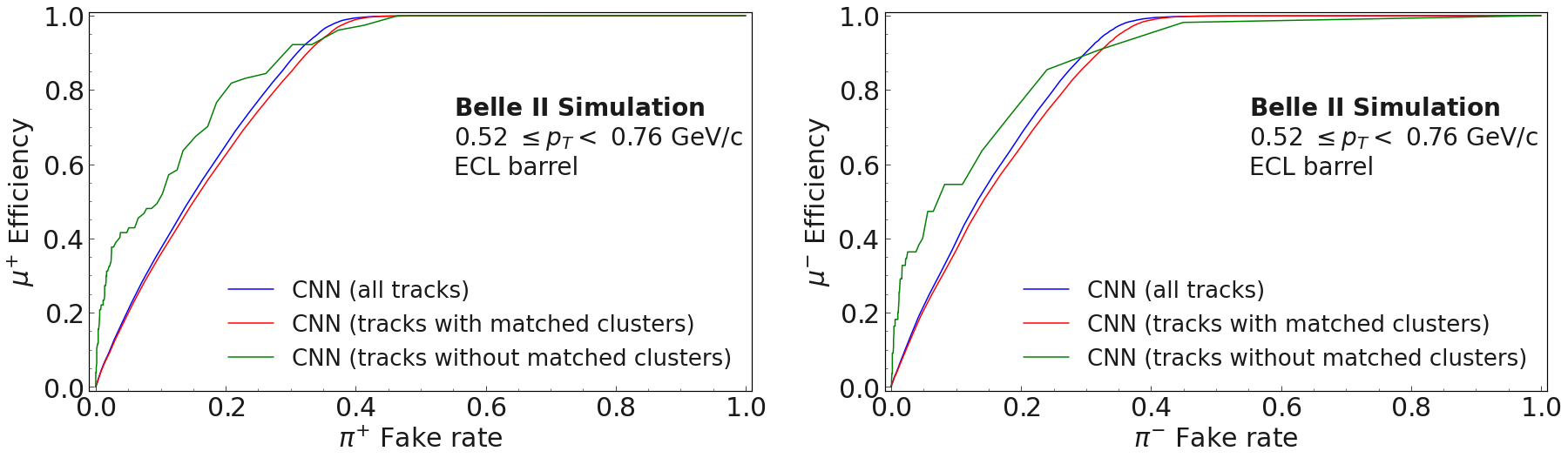}
	\caption{Comparison of CNN performance for all tracks, tracks with and without matched cluster. The left and right plots show positive and negative charged tracks, respectively.}
	\label{figs:roc_cnn_all}
\end{figure}

There are cases (mostly pions) that the software framework does not match a track with a cluster in the ECL. Since the CNN method does not rely on clustering, a comparison is made among all tracks and tracks with and without matched cluster in the inference phase. The comparison is shown in figure \ref{figs:roc_cnn_all} in which the blue ROC curves, representing the CNN (all tracks), can be considered as an average between the red and green curves, which represent tracks with and without matched clusters, respectively. The large difference between red and green ones is due to statistics. The green ROC curve is roughly 2.8\% and 3.5\% of the test dataset in the p$_{T}$ range of 0.52 -- 0.76 GeV/$c$ for positive and negative charged tracks, respectively.

\begin{figure} [htbp]
\centering
	\includegraphics[width=130mm]{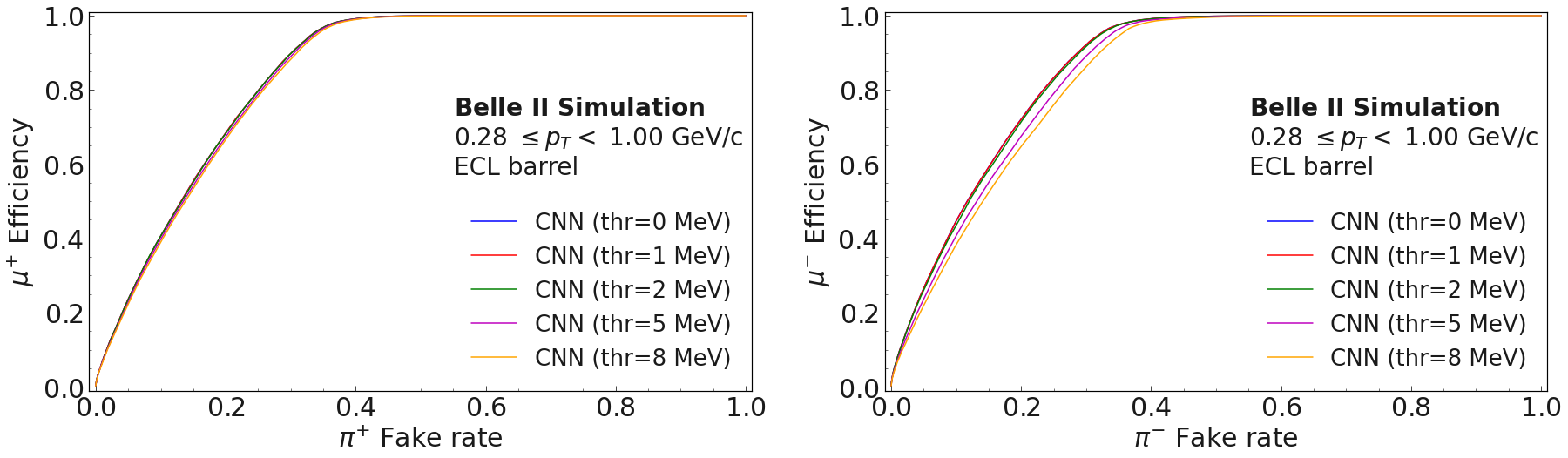}
	\caption{Comparison of CNN performance for different thresholds. The left and right plots show positive and negative charged tracks, respectively.}
	\label{figs:roc_cnn_thr}
\end{figure}

Due to the higher beam background level at Belle II, a higher minimal energy threshold for crystals may reduce the pile-up from beam background. In order to check the robustness of the CNN method against different levels of beam background, different energy thresholds for crystals of 0, 1, 2, 5, and 8 MeV are tested and the results are shown in figure \ref{figs:roc_cnn_thr} (zero means no threshold). Results show the CNN is robust against the choice of different energy thresholds.

\section{Summary and outlook}
This study shows that by using patterns of energy depositions in the ECL, muon-pion separation can be improved for low-momenta tracks. The pion fake rate is 4.2\% and 6.7\% larger at a typical working point of 90\% muon identification efficiency for positive and negative charged tracks, respectively. The CNN method is available in the Belle II analysis software framework and will be integrated as part of the standard Belle II reconstruction software. This study can be extended to include additional low-level ECL crystal information, e.g., pulse-shape discrimination \cite{Savino, PSD} which is useful to separate hadronic and electromagnetic interactions. In order to validate the CNN method on data, clean samples of muons and pions are selected using $e^{+} e^{-} \rightarrow \mu^{+} \mu^{-} \gamma$ and $D^{*\,+} \rightarrow D^{0}\,[ \rightarrow K^{+}\,\pi^{-}]\,\pi^{+}$, respectively. These results are underway.

\end{document}